\documentclass[prb,twocolumn,superscriptaddress,showpacs,aps,10pt]{revtex4-1}
\usepackage{color}
\usepackage{amsmath}
\usepackage{graphicx}
\usepackage{amssymb}
\usepackage{amsthm, amscd}
\usepackage{amsfonts}
\usepackage{times}
\usepackage{wasysym}

\usepackage[english]{babel}
\usepackage[utf8x]{inputenc}

\begin{document}

\title{Landau Level Mixing in the Perturbative Limit}

\author{Steven H. Simon}
\affiliation{Rudolf Peierls Centre for Theoretical Physics, University of Oxford, OX1 3NP, United Kingdom}

\author{Edward H. Rezayi}
\affiliation{Department of Physics, California State University, Los Angeles, California 90032}

\begin{abstract}
We study the effects of Landau level mixing in the limit of weak electron interaction.  We use a numerical method to obtain the two- and three-body corrections to quantum Hall pseudopotentials, which are exact to lowest order in the Landau level mixing parameter.  Our results are in general agreement with certain analytic results (some derived here, some derived by other authors) in the thermodynamic limit.    We find that the convergence to this thermodynamic limit can be slow.   This suggests that errors could occur if one tries to use pseudopotentials derived in a thermodynamic limit for numerical work on finite systems. 
\end{abstract}

\maketitle

\section{Introduction} 

Fractional quantum Hall effect is typically studied in the limit where the cyclotron   energy $\hbar \omega_c$ is large, so one can focus attention on only a single Landau level (LL).  However, in typical experiments, the interaction energy scale $E_{coulomb} = e^2/(\epsilon \ell_0)$, with $\ell_0$ the magnetic length, is often on the same order as the cyclotron energy, thus making Landau level mixing an important consideration. Historically, the difficulty of including LL mixing into calculations, along with the general belief that LL mixing would change predictions quantitatively, but not qualitatively, has generally prevented researchers from addressing this issue in much detail.   Relatively recently, however, it became clear that one could not always ignore LL mixing so easily.   Crucially, it was pointed out that LL mixing would be {\it required} in order to break the symmetry between two possible (nonabelian) quantum Hall states that might be realized at $\nu=5/2$ --- the Moore-Read\cite{MooreRead} state, and the so-called AntiPfaffian\cite{AntiPfaffian}.  In this context, the current authors used a Landau-level truncation scheme to conclude that in typical samples the Anti-Pfaffian is favored\cite{RezayiSimon}.

Although the experiments are not usually in the limit of small LL mixing, it is nonetheless natural to consider the case where LL mixing is weak. To this end, Bishara and Nayak\cite{Bishara} used a systematic perturbative approach to obtain corrections to leading order in the parameter $\kappa = E_{coulomb}/\hbar \omega_c$.    As pointed out in that work, the effect at leading order can be summarized as being a correction to the two-body\cite{Haldane} and three-body\cite{SimonRezayiCooper,Davenport} pseudopotentials.   These corrections were used in an exact diagonalization study in Ref.~\onlinecite{Wojs}.   

Almost two decades earlier, a seminumerical LL mixing study was undertaken by Rezayi and Haldane\cite{RezayiHaldane} to obtain the leading perturbative corrections to the two-body interaction.  Aware that the three-body interactions would be crucial, in the process of writing Ref.~\onlinecite{RezayiSimon}, the current authors explored extending the numerical technique of Ref.~\onlinecite{RezayiHaldane} to obtain three-body corrections as well --- the results of which will be presented in the current paper  (these results also update the results of Ref.~\onlinecite{RezayiHaldane}, which were necessarily rough given the limited computational ability of the time).    Unfortunately, we found that at least some of our results disagreed with the published results of Bishara and Nayak\cite{Bishara}.  We later supported our numerical results by an analytical calculation valid in the lowest LL, which we present in the appendix of this paper. 

In discussions with Nayak as well as with Matthias Troyer, it was suggested that the disagreement in results was due to a normal-ordering error in Ref.~\onlinecite{Bishara}.  (This is now discussed in some detail in Ref.~\onlinecite{Peterson}).   However, even fixing this error it was not obvious that the analytical results entirely agreed with the numerical calculation. 

More recently another fully analytical calculation was performed by Sodemann and Macdonald\cite{Sodemann}.  While the results agreed with our numerical calculation in the lowest LL (LLL), some disagreements remained in the first exicted LL (LL1). 

Still more recently, Peterson and Nayak have produced a new manuscript\cite{Peterson} performing an analytical calculation which now agrees with our numerical results  (their manuscript also discusses finite thickness corrections and graphene, as well as the conventional pure Coulomb interaction).    Since these results are controversial, and since it is obviously very easy to obtain incorrect results, we have decided to publish our numerical results to support the conclusions of Peterson and Nayak.  In so doing, we will also point out a key potential limitation of the analytic techniques of these references --- that they apply only in the thermodynamic limit. 

\section{Method}   

The method roughly follows that of Ref.~\onlinecite{RezayiHaldane}.  We work on a finite sphere with a monopole of strength $N_\phi$ flux.   We start with some number of completely filled Landau levels. We would like to consider the interactions of $n=2$ or 3 additional electrons added to the next empty LL, which we call the {\it valence} LL.   For the two (three) particle case we consider a basis of states with two (three) electrons added to the valence LL.    For simplicity we consider only states with total $L_z = 0$, which will be sufficient for our purposes.  One can diagonalize the $L^2$ operator within this space to pick out states with angular momentum $L$ which we call $|L,n\rangle$ where $n=2,3$ indicates whether we are considering two or three electrons added to the valence LL.   While specifying $L$ picks out a unique wavefunction for the case of $n=2$ electrons added to the valence LL, for the case of $n=3$ electrons added to the valence LL, the quantum number $L$ alone does not uniquely specify the state for $L=9$ and $L \geq 11$, in the spin polarized case, and for $L \geq 4$ in the unpolarized case\cite{SimonRezayiCooper,Davenport}.   In these cases, we need additional quantum numbers to specify a specific wavefunction.  Although we will not delve into such complexity in the current paper, the described technique can easily be extended to handle these cases as well. 

Our Hamiltonian can be abbreviated as 
$$
  H = \hat K + \hat U
$$
with $\hat K$ the kinetic terms and $\hat U$ the interaction term which we assume to be purely two-body  (and below we will take this to be of the Coulomb form).  The above mentioned $|L,n\rangle$ states all have the same kinetic energy $K^0_{n}$
$$
  K^0_{n} |L,n\rangle =\hat K |L,n\rangle
$$
 
We next apply the interaction to our basis of states 
$$
 \widetilde{| L,n\rangle} = \hat U|L,n\rangle  
$$
It should be noted that these wavefunctions contain components with electrons above the valence LL, as well as holes below the valence LL.   Crucially, for numerical work we must truncate the Hilbert space that we will consider by not allowing electrons to occupy Landau levels above some level $LL_{max}$.     Note that written in the standard single electron orbital basis, the number of components contained in a one of these wavefunctions $\widetilde{| L, n\rangle}$ is only polynomially large in $N_\phi$ and in $LL_{max}$.  

Note that the Coulomb energy of these wavefunctions without considering LL mixing can be written as
$$
  V^0_{L,n} = \langle L,n | \hat U | L,n \rangle  = \langle L,n \widetilde{|  L,n \rangle}.
$$
For the case of $n=2$ electrons in a valence LL, these values are precisely the pseudopotentials in the absence of LL mixing. They are proportional to $E_{coulomb}$. 

The leading LL mixing correction to the energy of the $|L\rangle$ state is then given by 
$$   
\delta V_{L,n} = \widetilde{\langle L,n |} \, \hat P\frac{1}{ K^0_n-\hat K}    \hat P \, \widetilde{|L,n \rangle}
$$
where $\hat P$ projects to the space where $\hat K \neq K^0_n$.    In the case of $n=2$ electron $\delta V_{L,2}$ is precisely the leading LL mixing correction to the $V^0_{L,2}$  pseudopotentials.   These energies are proportional to $\kappa E_{coulomb} = (e^2/\epsilon \ell_0)^2 / \hbar \omega_c$.  Note that in the {\it Results} section below we will always quote energies in units of $\kappa E_{coulomb}$.

For the $n=3$ electron case, we want to find the ``irreducible" part of the three-body interaction --- that is, the part of the energy shift that cannot be accounted for by a correction to the two-body pseudopotential.  We first calcualte $V^0_{L,2}$ and $\delta V_{L,2}$.  We then consider a modified Coulomb interaction within the single valence LL that consists only of $V^0_{L,2} + \delta V_{L,2}$ and contains no LL mixing terms.   We then evaluate the shift in     $V^0_{L,3}$ resulting from having added $\delta V_{L,2}$ to the bare pseudopotential $V^0_{L,2}$.   This shift is subtracted from $\delta V_{L,3}$ to give the irreducible three body pseudopotential.   Note that obtaining the three-body term thus requires obtaining the two-body term first.   

The described technique determines the leading order LL-mixing corrections to the interaction.  However, we must recall that we have truncated the Hilbert space to a finite number of LL's and we must numerically extrapolate to infinite $LL_{max}$.  Finally, since we are interested in comparing to results derived for large (planar) systems, we must also extrapolate to the limit of large $N_{\phi}$ corresponding to a large sphere.  Thus, the numerical results presented below are the results of a double extrapolation.  Note however if pseudopotentials are to be used on finite sized systems, for consistency, it is preferable to use pseudopotentials explicitly derived for the system size in question.  We will discuss this issue further below. 

\begin{figure}
\includegraphics[width=3in]{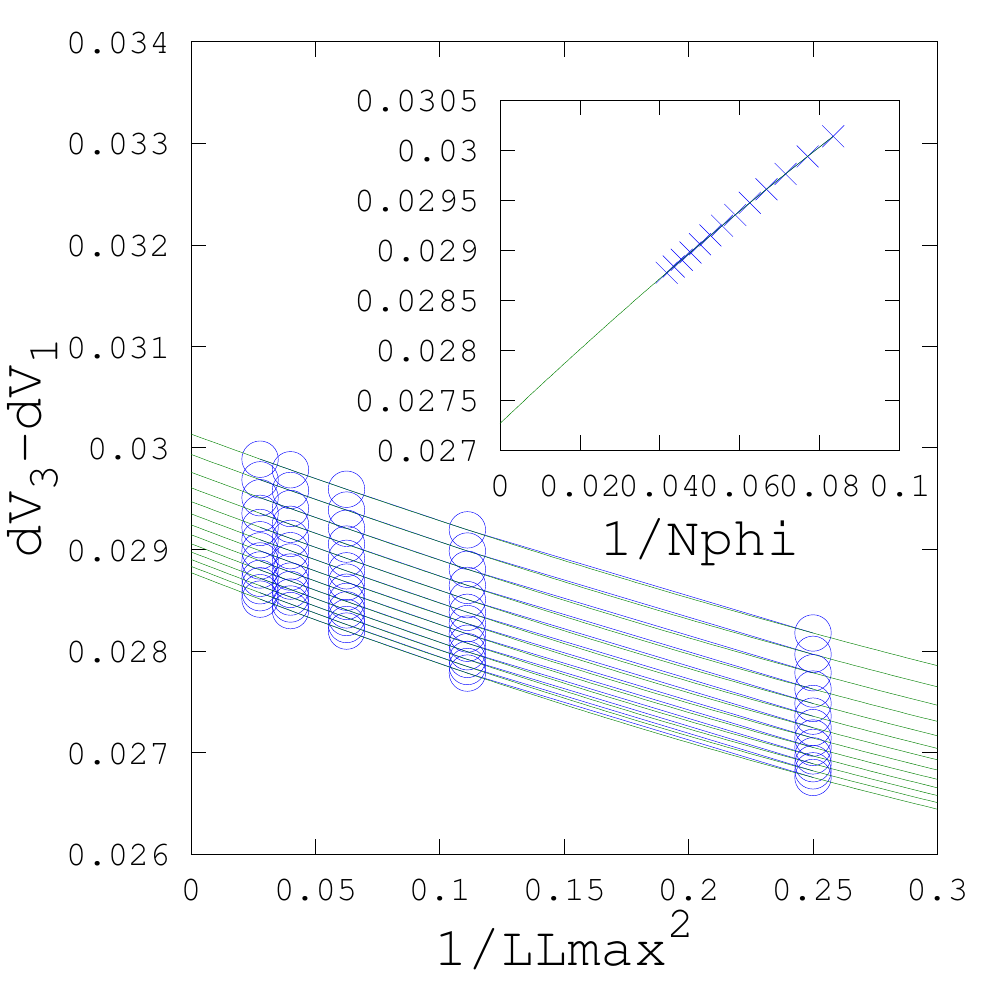}
\includegraphics[width=3in]{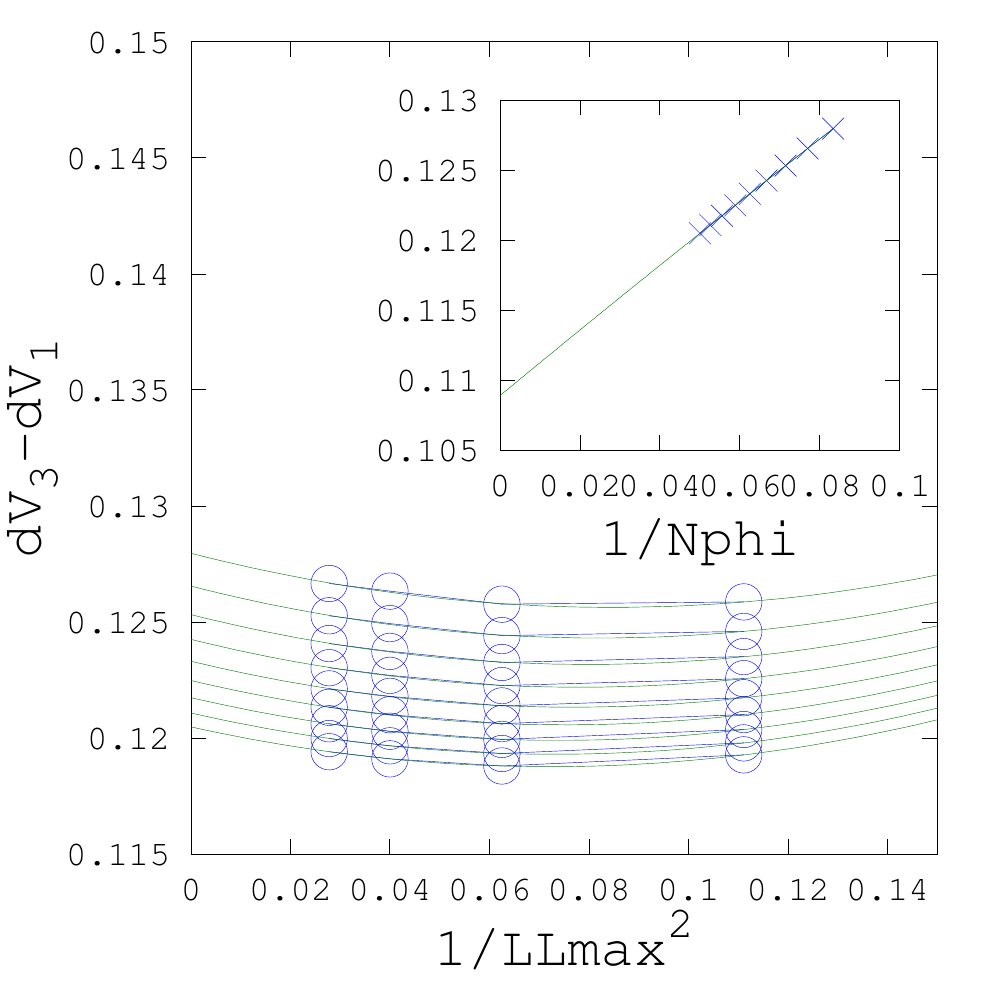}
\caption{(color online) Sample Double Extrapolations. The multiple curves in the main plots correspond to different $N_\phi$.   The intercepts (infinite $LL_{max}$) are then plotted in the inset and extrapolatCd to  infinite $N_\phi$.    {\bf Top:} Extrapolation of the two body LL-mixing correction terms $\delta V_3 - \delta V_1$ for electrons in the LLL.   Here $LL_{max}$ ranges from 2 to 6.  $N_{\phi}$ ranges from 12 to 24.  
The double extrapolation yields a value 0.0273. compared to the analytically exact result 0.0272267  (See Table I,  Refs.\onlinecite{Peterson,Sodemann}, and the appendix below in this paper).   {\bf Bottom:}  Extrapolation of the two body LL-mixing correction terms $\delta V_3 - \delta V_1$ for electrons in the 1st excited LL.   Here $LL_{max}$ ranges from 3 to 6.  $N_{\phi}$ ranges from 12 to 20.   Due to the curvature in these plots, one must assume a bit more error on the extrapolation.  The double extrapolation yields a value 0.1089.    The analytical reslult of reference \onlinecite{Peterson} yields .1104.  This seems well within the range of possible extrapolation error.}
\label{fig:plots}
\end{figure}

\section{Results}  

All numerical results will be quoted in units of $\kappa E_{coulomb}$. To avoid having to worry about electron self-energies (which may converge more slowly) it is more convenient to keep track of differences in pseudopotential coefficients.  This is natural given that adding a common constant to all pseudopotential coefficients gives an overall shift in energy but leaves the eigenstates of the Hamiltonian unchanged.   To calibrate our technique, we consider the two body pseudopotential in the lowest LL (LLL) as shown in the top of Fig.~\ref{fig:plots}.   Analytic results are derived for this case in the appendix (see also Refs.~\onlinecite{Peterson,Sodemann}).  In this case, the numerical extrapolation is fairly clean.  However, in the case of the two-body corrections in the first excited LL (LL1), as shown in the bottom of Fig.~\ref{fig:plots}, the extrapolation involves a guess at the proper extrapolation functional form; as a result one should assume a bit more error in the result.  Nonetheless, comparing our result with the analytic prediction of Ref.~\onlinecite{Peterson} still obtains agreement within about 1\%.

\begin{table}
\begin{tabular}{c|c|c}
    &   Numerical &  Analtyic.   Appendix   \\
LLL & This work &    of this work and \onlinecite{Peterson,Sodemann} \\
 \hline
$\delta V_3 - \delta V_1$  & .0273 &   .0272267  \\
$\delta  V_5 - \delta V_1$  & .0306 &   .0305804   \\
$\delta  V_7 - \delta V_1$  & .0316 &   .0316038  \\
$\delta V_9 - \delta V_1$  & .0321 &   .0320453    \\
\end{tabular}

\vspace*{15pt}

\begin{tabular}{c|c|c|c}
    & \multicolumn{2}{ c|}{Numerical} &  \\
LL1 & This work & Ref.~\onlinecite{RezayiHaldane} & Analtyic \onlinecite{Peterson} \\
\hline
$\delta V_3 - \delta V_1$  & .109 & .1095  &  .1104   \\
$\delta V_5 - \delta V_1$  & .175 & .1688  &  .1790   \\
$\delta V_7 - \delta V_1$  & .198 & .1849  &  .2028  \\
$\delta V_9 - \delta V_1$  & .206 &        &  .2120  \\
\end{tabular}
\caption{Predictions for the two-body pseudopotential leading corrections due to LL mixing for electrons in a spin polarized LL.    The top table is results for the LLL, the bottom table is the results for LL1.  Again, we quote differences in pseudopotentials to remove physically unimportant self-energy terms.    As suggested by the bottom panel of Fig.~\ref{fig:plots}, in the case of LL1, one should assume extrapolation error of possibly as much as .005.  As in that figure the extrapolation considers  $LL_{max} \leq 6$ and $N_{\phi} \leq 20$.}
\label{tab:twobody}
\end{table}

In Table \ref{tab:twobody} we show LL mixing corrections to the two-body pseudopotentials in the LLL and in LL1.   In the LLL the agreement between the numerics and the analtyics is excellent.  In LL1, the agreement is less precise, but considering the difficulty in extrapolation (See Fig.~\ref{fig:plots}), the agreement can still be considered to be fairly good.

\begin{table}
\begin{tabular}{c|c|c}
LLL & Numerical & Analytic\cite{Peterson,Sodemann} \\
  \hline
$V_5 - V_3$  & .0214  &  .0214 \\
$V_6 - V_3$  & .0060  &  .0074 \\
$V_7 - V_3$  & .0239  &  .0240 \\
$V_8 - V_3$  & .0123  &  .0133 \\
\end{tabular}

\vspace*{15pt}

\begin{tabular}{c|c|c}
LL1 & Numerical & Analytic\cite{Peterson,Sodemann} \\
 \hline
$V_5 - V_3$  & .0109  &  .0093 \\
$V_6 - V_3$  & .0027  &  .0048 \\
$V_7 - V_3$  & .0191  &  .0152 \\
$V_8 - V_3$  & .0177  &  .0138 \\
\end{tabular}
\caption{Predictions for the three-body pseudopotential coefficients due to LL mixing for electrons in a spin polarized LL.  Analytic results are taken from both Refs.~\onlinecite{Peterson} and \onlinecite{Sodemann}.  Again, we quote differences in pseudopotentials to remove physically unimportant self-energy terms.  The numerical results in the LLL (top) are extrapolate extremely accurately to the thermodynamic limit.   However,  for LL1 (bottom) the agreement is less good.  (See text for discussion). }
\label{tab:threebody}
\end{table}

\begin{figure}
\includegraphics[width=3in]{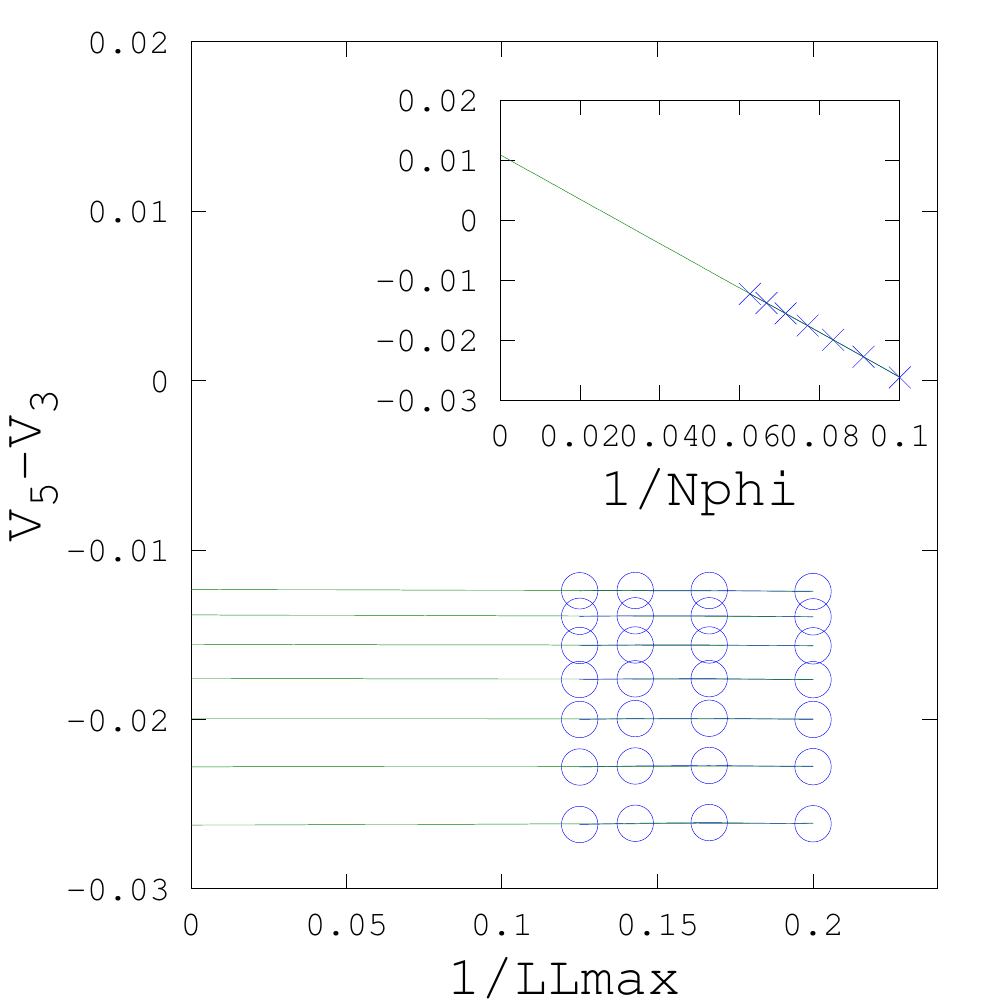}
\includegraphics[width=3in]{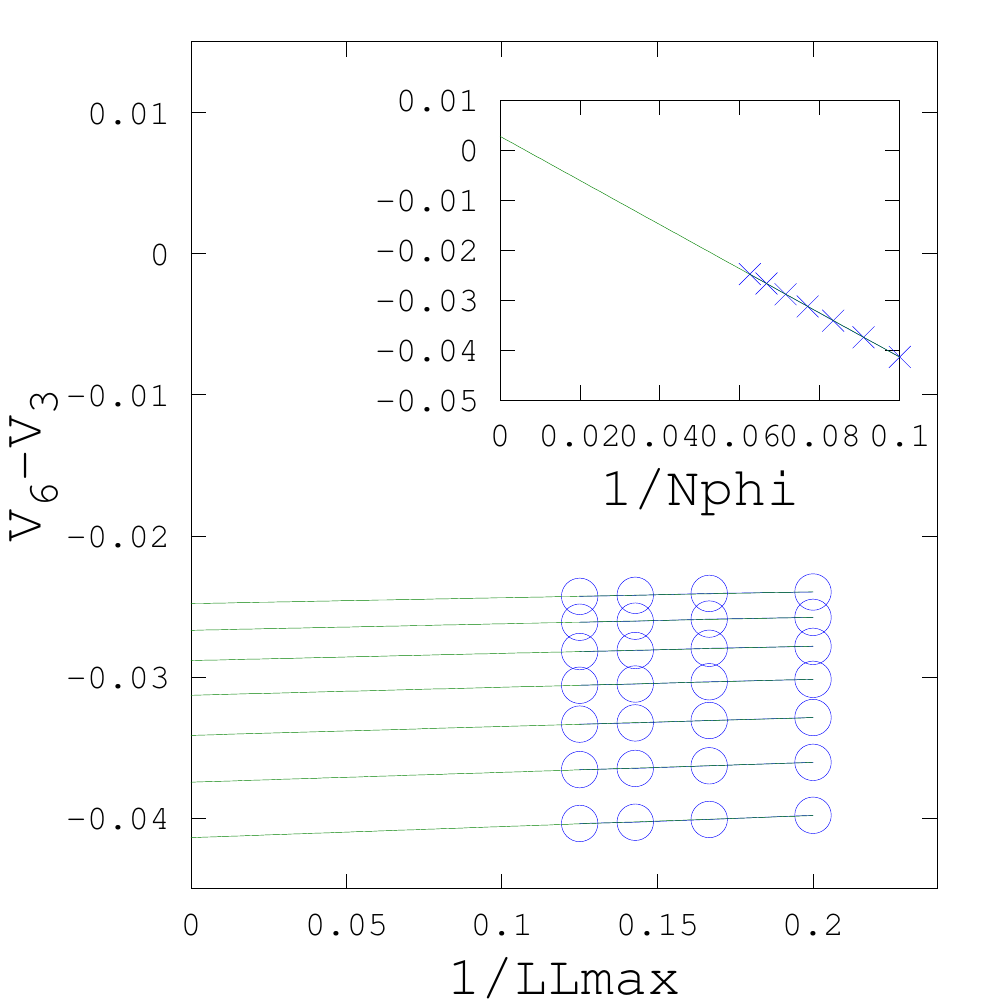}
\caption{(Color online) Sample double extrapolations of the three body pseudopotential $V_5 - V_3$ (top) and $V_6 - V_3$ (bottom) in LL1.  The extrapolated values are 0.109 and .0027 respectively.   Note that there is quite a bit of slope with respect to $N_\phi$ which makes accurate extrapolation difficult. Secondly note that the extrapolated result is close to zero on the scale of the values measured at finite $N_\phi$ which also makes the extrapolation difficult. }
\label{fig:plots2}
\end{figure}

In Table \ref{tab:threebody} we show the three-body pseudopotentials resulting from LL mixing for a spin polarized valence LL.   In the LLL the agreement between numerics (this work) and analytics (Refs.~\onlinecite{Peterson,Sodemann}) is quite good, whereas in LL1, the agreement is much less good.   To examine this situation a bit more closely, we show an example of some of the extrapolations in Fig.~\ref{fig:plots2}.   Note that there is quite a bit of change in the value of the pseudopotential as a function of $N_\phi$, which makes extrapolation difficult.  Nonetheless, given the uncertainties in the extrapolation, the agreement with the analytic results appears acceptable.

The fact that the convergence with system size is somewhat worse in LL1 than in LLL is to be expected being that all correlation lengths are larger in LL1 (the single particle wavefunctions are in some sense ``fatter").   In addition, working in LL1 is far more computationally demanding than in LLL since one needs to account for a far greater variety of excitations that involve exciting electrons out of the fully filled Landau levels (including both spin species).  This has limited our numerical calculations in LL1 to somewhat smaller systems.  While these particular calculations were performed on a not-very-powerful computer, the results are clear enough to demonstrate the general trends, and also to confirm agreement with analytic work by other authors.

\begin{figure}
\includegraphics[width=3in]{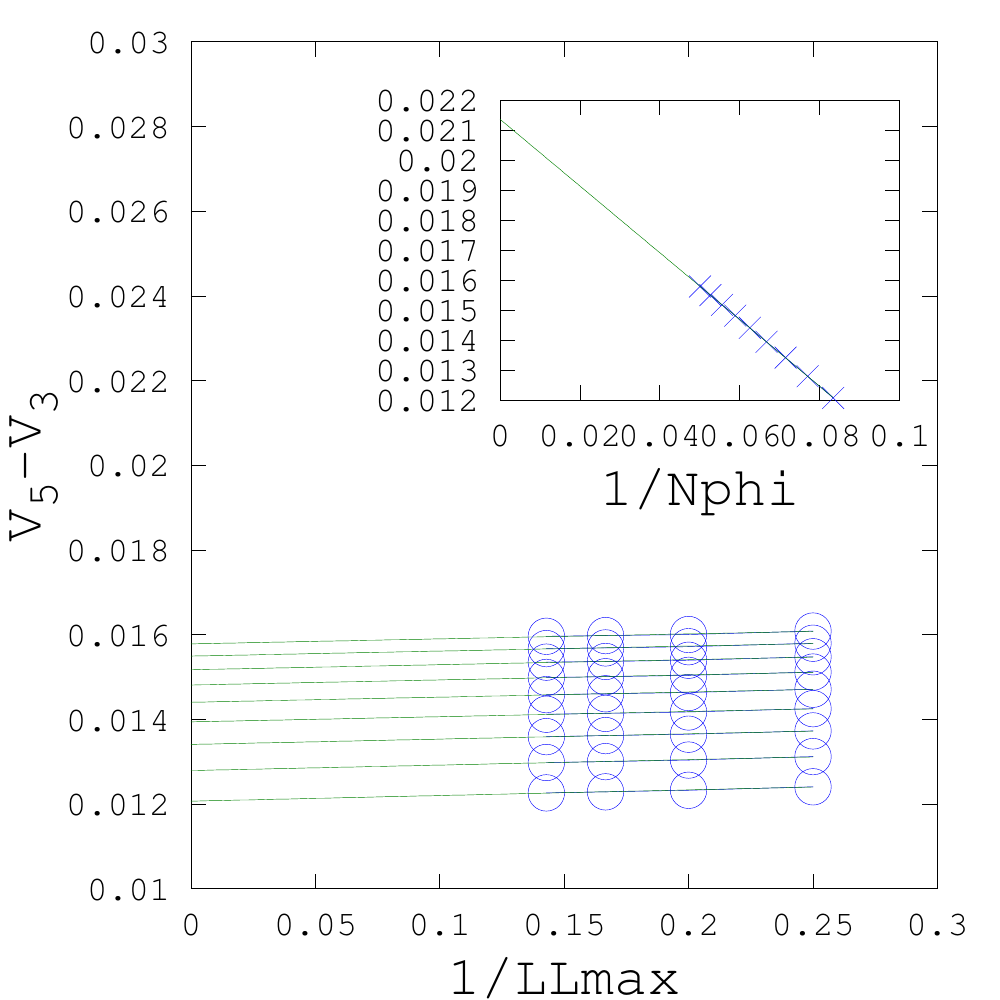}
\caption{(Color online) Sample double extrapolations of the three body pseudopotential $V_5 - V_3$ in the LLL.  The extrapolated values is .0214    Note that there the value of the pseudopotential taken at finite $N_\phi$ can be quite different from the thermodynamic limit.}
\label{fig:plots3}
\end{figure}

The slow convergence of the three-body potential raises a serious issue.   To a large extent, the purpose of deriving pseudopotentials is to use them for numerical exact diagonalizations for finite systems.  However, there is a clear inconsistency in using pseudopotentials derived in the thermodynamic limit for exact diagonalizations of a finite sized system.  While the size dependence of the two-body pseudopotentials is relatively minor, the size dependence of the three-body term appears to be very significant.  This is also true for the LLL as well (perhaps to a slightly lesser extent), as shown in Fig.~\ref{fig:plots3}. This raises the possibility that, for cases where the three-body potential is important in determining the physics, one will not see the true thermodynamic behavior until very large system size.   We suggest that finite size extrapolations may have fewer errors if diagonalizations of a given size system use pseudopotentials designed for that particular size. 

To elaborate on the issue of extrapolation, we recall the work of Ref.~\onlinecite{Morf}.  There it was pointed out that extrapolation converges rapidly if (a) one uses interaction  (such as chord distance) appropriate for each particular size system and (b) one rescales the interaction strength to keep the charge density constant. For this reason, in the history of quantum Hall numerics, the vast majority of calculations have used pseudopotentials derived for each particular system size in question.  While we cannot say with certainty that other schemes may not be made to work, it seems very risky to throw out the inherent consistency of this established approach.

\section{Summary}
The LL mixing corrections presented here should be of value in the analysis of physical systems where LL mixing terms are weak (very high density quantum Hall samples). Moreover, the results here have been used to independently establish the validity of analytical results of Ref.~\onlinecite{Peterson} and some of the results of Ref.~\onlinecite{Sodemann}.   Finally, we point out a potential inconsistency in using pseudopotentials derived in the thermodynamic limit for calculations on finite systems. 

After the initial submission of this paper, a revised version of Ref.~\onlinecite{Sodemann} appeared which is in full agreement with this paper as well as with the results of Ref.~\onlinecite{Peterson}.

{\bf Acknowledgments:} The authors are grateful to Arek Wojs, Jainendra Jain, Inti Sodemann, Allan MacDonald, Chetan Nayak, Michael Peterson, and Matthias Troyer.   We thank Arek Wojs for sharing his numerical LL truncation results with us.   We thank Matthias Troyer for helping to identify normal ordering as a crucial issue.   We thank Inti Sodemann and Allan Macdonald for convincing us that, to lowest order, the calculation of two and three particles in a given LL is complete information about LL mixing independent of the filling of the LL. SHS acknowledges funding from EPSRC grants EP/I032487/1 and EP/I031014/1.  SHS also acknowledges the Aspen Center for Physics where some of the key discussions took place.  EHR acknowledges DOE Grant No. DE-SC0002140.

{\bf Appendix}   In this appendix we analytically calculate the leading LL mixing correction to the two-body pseudopotential in the lowest LL. 

For a single electron in a magnetic field define\cite{LLalgebra}
\begin{align}
 a &= \sqrt{2} \left( \ell \bar \partial_z + \frac{1}{4 \ell} z \right) \\
 b &= \sqrt{2} \left( \ell \partial_z + \frac{1}{4 \ell} \bar z \right)
\end{align}
Where $\ell$ is the magnetic length. Noting that $\partial_z^\dagger = -\bar \partial_z$ and $z^\dagger = \bar z$ we have
\begin{equation}
 [a, a^\dagger] =  [b, b^\dagger] =1
\end{equation}
The Hamiltonian is
\begin{equation}
 H = \hbar \omega_c (a^\dagger a + \frac{1}{2})
\end{equation}
and the z-component of angular momentum is
\begin{equation}
 L_z = \hbar \left( b^\dagger b - a^\dagger a  \right)
\end{equation}
The single particle eigenstates with energy $(n + \frac{1}{2}) \hbar \omega_c$ and $L_z = \hbar (m - n)$ can be generated as
\begin{equation}
 |n,m\rangle = \frac{(b^\dagger)^m (a^\dagger)^n}{\sqrt{n! m!}} |0 0 \rangle
\end{equation}
where $|0, 0 \rangle$ is the fiducial state for which $a|0 0\rangle = b|0 0\rangle = 0$, which takes the form
\begin{equation}
 \langle \vec r | 0 ,0 \rangle = \frac{1}{\sqrt{2 \pi \ell^2}} e^{-|z|^2/4 \ell^2}
\end{equation}
The explicit form of all the single particle eigenstates can be shown to be
\begin{align}
& \langle \vec r | n, m \rangle =  \\
&(-1)^n \sqrt{\frac{n!}{2 \pi \ell^2  m!}} \left( \frac{z}{\ell \sqrt{2}} \right)^{m-n} L_n^{m-n}\left( \frac{|z|^2}{2 \ell^2} \right)e^{-|z|^2/4 \ell^2}
\end{align}

Now consider two electrons with coordinates $z_1$ and $z_2$.   We can transform to center of mass $z_C$ and relative coordinates $z_R$
\begin{equation}
 \left( \begin{array}{c} z_R \\ z_C \end{array} \right) =  \left( \begin{array}{cc}1 & -1 \\ \frac{1}{2} & \frac{1}{2} \end{array} \right)
 \left( \begin{array}{c} z_1\\ z_2 \end{array} \right)
\end{equation}
This transformation has unit Jacobian.  Correspondingly we have
\begin{equation}
 \left( \begin{array}{c} \partial_{z_R} \\ \partial_{z_C} \end{array} \right) =  \left( \begin{array}{cc} \frac{1}{2} & -\frac{1}{2} \\ 1 & 1 \end{array} \right)
 \left( \begin{array}{c} \partial_{z_1}\\ \partial_{z_2} \end{array} \right)
\end{equation}
which maintains the desired commutations $[\partial_{z_R}, z_{R}] =[\partial_{z_C}, z_{C}] =  1$.

It is convenient to write out the reverse transformations as well
\begin{equation}
 \left( \begin{array}{c} z_1\\ z_2 \end{array} \right)
 =  \left( \begin{array}{cc}\frac{1}{2} & 1 \\ -\frac{1}{2} & 1 \end{array} \right) \left( \begin{array}{c} z_R \\ z_C \end{array} \right)
\end{equation}
and
\begin{equation}
\left( \begin{array}{c} \partial_{z_1}\\ \partial_{z_2} \end{array} \right)
=  \left( \begin{array}{cc} 1 & \frac{1}{2} \\ -1 & \frac{1}{2} \end{array} \right)
    \left( \begin{array}{c} \partial_{z_R} \\ \partial_{z_C} \end{array} \right)
\end{equation}
Substituting this into our expression for $a_i$ we obtain
\begin{equation}
a_{1,2} = \sqrt{2}\left[\left(\frac{\ell}{2} \bar \partial_{z_C} + \frac{1}{4 \ell} z_C \right) \pm  \left( \ell \bar \partial_{z_R} + \frac{1}{8 \ell} z_R  \right) \right]
\end{equation}
Defining
$  \ell_R = \ell \sqrt{2} $ and  $\ell_C = \ell / \sqrt{2}$ we can rewrite this as
\begin{align} \nonumber
a_{1,2} &= \frac{1}{\sqrt{2}} \left[\sqrt{2}\left(\ell_C \bar \partial_{z_C}  + \frac{1}{4 \ell_C} z_C \right) \right. \\ & ~~~~~~\pm \left. \sqrt{2}\left(\ell_R \bar \partial_{z_R} + \frac{1}{4 \ell_R} z_R   \right) \right] \\
  & = \frac{1}{\sqrt{2}}(a_C \pm a_R)
\end{align}
where we have defined the operators
\begin{align}
a_{R} &= \sqrt{2}\left(\ell_R \bar \partial_{z_R} + \frac{1}{4 \ell_R} z_R  \right)
\\
a_C & =
 \sqrt{2}\left(\ell_C \bar \partial_{z_C} + \frac{1}{4 \ell_C} z_C \right)
\end{align}
which correctly satisfy $[a_R, a_R^\dagger] = [a_C, a_C^\dagger] = 1$.

Similarly we may define operators
\begin{align}
b_{R} &= \sqrt{2}\left(\ell_R  \partial_{z_R} + \frac{1}{4 \ell_R} \bar z_R  \right)
\\
b_C & =
 \sqrt{2}\left(\ell_C  \partial_{z_C} + \frac{1}{4 \ell_C} \bar z_C \right)
\end{align}
such that
\begin{equation}
b_{1,2} = \frac{1}{\sqrt{2}}(b_C \pm b_R)
\end{equation}
and again $[b_R, b_R^\dagger] = [b_C, b_C^\dagger] = 1$.

In terms of these new variables, we have the Hamiltonian
\begin{align}
 H &= \hbar \omega_c \left( a_1^\dagger a_1 + a_2^\dagger a_2 \right)  \\
 &= \hbar \omega_c \left( a_R^\dagger a_R + a_C^\dagger a_C\right)
\end{align}
And similarly the total z-component of angular momentum
\begin{align}
 L_z &= \hbar\left( b_R^\dagger b_R + b_C^\dagger b_C - a_R^\dagger a_R - a_C^\dagger a_C\right)
\end{align}

The key here is that a radial interaction $V(|r_1 - r_2|) = V(|z_R|)$ between the two particles cannot effect the center of mass degree of freedom, so we may freely choose all of the $a_C$ and $b_C$ oscillators to remain in their ground states.  Our kets will then be of the form $|N, M\rangle$ where $N$ and $M$ are the occupancies of the $a_R$ and $b_R$ oscillators.

Our initial state will have both electrons in the LLL and will give them relative angular momentum $M$.  Thus we choose an initial state  $|0, M\rangle$ for the $a_R$ and $b_R$ oscillators.  A rotationally symmetric interaction cannot change $L_z =b_R^\dagger b_R - a_R^\dagger a_R$, thus we can only have matrix elements of the form
\begin{equation}
  \langle N, M+N|V| 0, M  \rangle
\end{equation}
We can directly evaluate this matrix element as
\begin{widetext}
\begin{align}
& \int {\bf dz_R}  (-1)^N \sqrt{\frac{N!}{2 \pi \ell^2  (M+N)!}} \left( \frac{\bar z_R}{\ell_R \sqrt{2}} \right)^{M} L_N^{M}\left( \frac{|z_R|^2}{2 \ell_R^2} \right)e^{-|z_R|^2/4 \ell_R^2}\,\,\, V(|z|) \,\, \sqrt{\frac{1}{2 \pi \ell_R^2  (M)!}} \left( \frac{z_R}{\ell_R \sqrt{2}} \right)^{M} e^{-|z_R|^2/4 \ell^2} \\
=& \frac{1}{\ell_R^2} \sqrt{\frac{N!}{M! (M+N)!}} \int_0^\infty r dr  V(r)  \left(\frac{r^2}{2 \ell_R^2} \right)^M  L_N^{M}\left( \frac{r^2}{2 \ell_R^2} \right) e^{-r^2/2 \ell_R^2}
\end{align}
Let us now plug in the coulomb form of the interaction
$
  V = e^2/(\epsilon r)
$
and making the substitution $x=r^2/2 \ell_R^2$ we obtain
\begin{equation}
  \langle N, M+N|V| 0, M  \rangle = \frac{e^2}{\epsilon \ell_R} \frac{1}{\sqrt{2}} \sqrt{\frac{N!}{M! (M+N)!}} \int_0^\infty \, dx  \, x^{M-1/2} \, L_N^M(x)\, e^{-x}
\end{equation}
\end{widetext}

The integral is standard on integral tables and can be performed to give
\begin{equation}
  \langle N, M+N|V| 0, M  \rangle = \frac{e^2}{\epsilon \ell} \frac{\Gamma(M+1/2)\Gamma(N+1/2)}{2 \sqrt{ \pi \, N! \, M! \, (N+M)! }}
\end{equation}
and note that this is now written in terms of the original magnetic length.

As a check we can determine the Coulomb pseudopotentials in the LLL with no LLL mixing. These would be given by
\begin{equation}
V_M =  \langle 0, M |V| 0, M  \rangle = \frac{e^2}{\epsilon \ell} \frac{\Gamma(M+1/2)}{  M! 2}
\end{equation}
which matches the well known expression\cite{Yoshioka}.

Now in second order perturbation theory the change in energy of the $|0 , M\rangle$ state should be given by
\begin{equation}
\delta V_M =  \sum_{N > 0}  \frac{|  \langle N, M+N|V| 0, M  \rangle |^2}{-\hbar \omega_c N}
\end{equation}

This result agrees with the analytic expression given in Refs.~\onlinecite{Peterson,Sodemann}.


\begin{thebibliography}{50}



\bibitem{MooreRead}  G. Moore and N. Read, Nucl. Phys. {\bf B360},362 (1991).

\bibitem{AntiPfaffian} M.~Levin, B.~I.~Halperin and B.~Rosenow, Phys.~Rev.~Lett.~{\bf 99}, 236806 (2007); S.-S.~Lee, S.~Ryu, C.~Nayak and M.~P.~A.~Fisher, Phys.~Rev.~Lett. {\bf 99}, 236807 (2007).

\bibitem{RezayiSimon} E.~H. Rezayi and S.~H. Simon, Phys.~Rev.~Lett.~{\bf 106}, 116801 (2011).


\bibitem{Bishara}  W.~Bishara and C.~Nayak, Phys.~Rev.~{\bf B80}, 121302(R)
(2009).


\bibitem{Haldane} F.~D.~M.~Haldane, Phys.~Rev.~Lett.~{\bf 51}, 605 (1983).

\bibitem{SimonRezayiCooper} S.~H.~Simon,  E.~H.~Rezayi, and N.~R.~Cooper, Phys.~Rev.~{\bf B75}, 195306 (2007).

\bibitem{Davenport} S.~C.~Davenport and S.~H.~Simon 
Phys.~Rev.~{\bf B85}, 075430 (2012).


\bibitem{Wojs} A.~Wojs, C.~Toke and J.~K.~Jain,  Phys.~Rev.~Lett.~{\bf 105}, 096802 (2010)


\bibitem{RezayiHaldane}  E.~H.~Rezayi and F.~D.~M.~Haldane, Phys.~Rev.~{\bf B42}, 4532 (1990).


\bibitem{Peterson} M.~R.~Peterson and C.~Nayak,  arXiv:1303.1541. 

\bibitem{Sodemann} I.~Sodemann and A.~H.~MacDonald, arXiv:1302.3896.


\bibitem{Morf} R.~Morf and B.~I.~Halperin, Z.~Phys.~B~{\bf 68}, 391  (1987). 



\bibitem{LLalgebra} For details of this raising and lowering operator formalism see for example the appendix of Kivelson et al, Phys.~Rev.~{\bf 36}, 1620, (1987).  Or see the book {\it The Quantum Hall Effects}, T.~Chakraborty and  P.~Pietiläinen, Springer-Verlag, (1995). 

\bibitem{Yoshioka} See for example, {\it The Quantum Hall Effect}, D. Yoshioka, Springer-Verlag (1998). 

\end{thebibliography}
\end{document}